\documentclass[journal]{IEEEtran}
\usepackage{graphicx,epsf}
\usepackage[usenames,dvipsnames,svgnames,table]{xcolor}

\usepackage{cite}

\ifCLASSINFOpdf
\else
\fi

\usepackage[cmex10]{amsmath}

\begin{document}
\title{Synchronization in arrays of vacuum microdiodes}

\author{Marjan~Ilkov,
        Kristinn~Torfason,
        Andrei Manolescu,
        \' Ag\' ust~Valfells
\thanks{Authors are with the School of Science and Engineering, Reykjavik University, 101 Reykjavik, Iceland}
\thanks{Manuscript received PUT DATE; revised PUT DATE.}}

\maketitle

\begin{abstract}
Simulations have shown that space-charge effects can lead to 
regular modulation of photoemitted beams in vacuum diodes 
with gap sizes on the order of 1 micron and accelerating voltage on 
the order of 1V. These modulations are in the THz regime 
and can be tuned by simply changing the emitter area or 
accelerating vacuum field. The average current in the diode 
corresponds to the Child-Langmuir current, but the amplitude 
of the oscillations is affected by various factors. Given the 
small size and voltage of the system, the maximum radiated 
AC power is expected to be small. In this work we show that 
an array of small emitters produces higher frequency signals 
than a single large emitter of same area, and how these 
emitters may be synchronized to produce higher power 
signals.

\end{abstract}


\IEEEpeerreviewmaketitle

\section{Introduction}

%
%
%
%
\IEEEPARstart{T}{erahertz} (THz) radiation is an active field of research with applications in communications, security screenings, molecular spectroscopy, medicine, and deep-space research to name a few examples \cite{siegel2002a} 
-\cite{federici2010}. Although there is no unique definition for the term Terahertz radiation it is commonly used to refer to the frequency range 300 GHz - 3 THz \cite{siegel2002a}.  In this range there is a lack of sources that can deliver appreciable power, particularly compact sources.  This is the so-called „Terahertz-gap“ \cite{booske2008}.
Among the most succesful THz sources are quantum-cascade lasers (QCL) \cite{scalari2009},\cite{williams2007} and vacuum electronic devices (VED) \cite{booske2008},\cite{booske2011},\cite{booske2005}. 
QCL have some limitations however.  They must be cryogenically cooled and are limited to producing 100‘s of milliwatts of THz radiation.  Representative parameters show a frequency range of 0.84 - 5.0 THz, a maximum operating temperature of 169 K for pulsed radiation and 117 K for CW, while maximum power is 250 mW for pulsed and 130 mW for CW radiation \cite{williams2007}.  VED have been able to produce considerable THz power \cite{booske2011}, particularly free-electron-lasers (FEL) \cite{freund1999} and gyrotrons \cite{nusinovich2004}, but also backward-wave-oscillators (BWO), klystrons and traveling-wave tubes (TWT) \cite{booske2011},\cite{booske2005}. 
 
Due to the inherent superiority of VED to solid state devices for producing high-power at high frequency \cite{booske2005},\cite{symons1998} it is natural to pursue that avenue in search of efficient, high-power THz sources.  However, it should be noted that the high power VED devices are both extremely large and expensive \cite{booske2008},\cite{booske2011}.  Nonetheless, with the advent of modern manufacturing techniques there is the promise of devising compact VED THz sources that are superior to solid state devices and QCL \cite{booske2011}, \cite{booske2005}, \cite{shin2009}
 -\cite{ives2004}.
 
Recent simulations of nanoscale vacuum diodes have indicated a mechanism for bunching of the beam from the cathode with a frequency corresponding to THz \cite{pedersen2011}.  The mechanism is based upon copious photoemission from a cold cathode, where the injected current is much greater than the space charge limit \cite{birdsall1966}
-\cite{caflisch2012}. Electrons are emitted from the cathode at a high rate until their density is such that they inhibit further emission.
As this bunch of electrons is accelerated away from the cathode the effect of its space-charge field at the cathode diminishes to a point where the orientation of the surface field at the cathode becomes favorable and emission resumes, resulting in the formation of a new bunch.  

For suitable diode dimensions (of the order of 1 $\mu$m), emitter area (scale length on the order of 100 nm), and potential difference applied to the diode (the order of 1V) it is possible to generate a continuous stream of electron bunches which arrive at the anode with intervals corresponding to THz frequency.  
The frequency is determined by the vacuum electric field in the diode, and the radius of the emitting area on the cathode \cite{jonsson2013}.  This mechanism is a many-electron version of the well known Coulomb blockade familiar in single electron transport in nanosystems.  The THz oscillation has, in fact, been shown to occur for Coulomb blockade in single electron emitters \cite{zhu2013}.

Since the beam modulation described in the preceding paragraph is persistent and easily tunable in the THz range, simply by varying the DC potential applied to the diode, it is tempting to examine the possibility of using it as a practical THz generator, either directly radiating or as a bunched electron source for a compact vacuum electronic amplifier.  However, the current from such a microdiode is typically around tens of microamperes for an applied potential around 1V \cite{jonsson2013}.  Thus, the expected power output from a single diode would be quite small.  Increasing the emitter area is not a satisfactory option to increase the output power as the bunching frequency decreases, and the quality of the bunches degrades with increasing emitter radius \cite{jonsson2013}.  From these considerations the idea sprang whether it might be possible to synchronize an array of emitters in order to generate  a coherent signal of increased power and THz frequency.  In general, synchronization means adjustment of rhythms in self-sustained periodic oscillators due to their weak interaction \cite{pikovsky2001}. If two oscillators with the same frequency synchronize, their instantaneous phase difference is zero.  This would lead to strengthening of the signal and increase power output.  However, the individual frequencies are expected to drop due to the interaction.
In the present work the interaction of electron bunches from emitter arrays is studied to look for evidence of synchronization and to understand the physical principle behind it.  This is done by using the same molecular dynamics approach as in previous research on microdiodes at Reykjavik University \cite{pedersen2011}, \cite{jonsson2013}.  In section \ref{System model} a brief description of the simulation methodology and model will be given.  In section \ref{Relative phase} the relative phase and interaction coefficient will be introduced.  Results will be presented in Section \ref{Results}, followed by a discussion and summary in section \ref{Discussion}.

\section{System model and simulation method}
\label{System model}
The system under consideration consists of a planar microdiode of infinite area, with photoemission taking place from prescribed areas on the cathode.  The number and configuration of the emitting areas can be varied, but the size of each emitter and average rate of photoemission for all of the emitters is the same.  This is shown schematically in figure 1.  The important parameters are the gap spacing of the diode, $D$, the potential applied to the diode, $\Phi$, the emitter radius, $R$, and the spacing between the center of adjacent emitters, $L$.  It is assumed that electrons are ejected from the emitter, via photoemission, at a rate which is much greater than the space-charge limiting current.  In other words, the current is never source-limited but always space-charge limited.  It is assumed that the emission velocity is negligible.  This space-charge limit is inherently guaranteed by the algorithm used in the simulation as will be described subsequently.

\begin{figure}
\begin{center}
\includegraphics [width=8.8 cm] {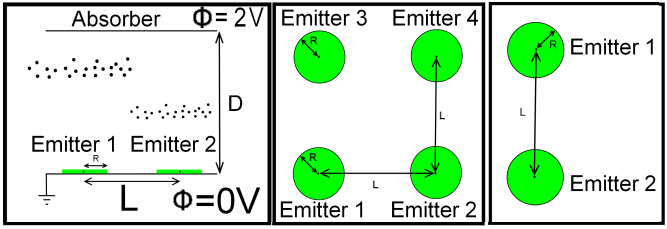}
\end{center}
\caption{Side view of the microdiode showing a cross section taken through the center of two emitters with black dots representing electrons (left). Top view of the cathode showing a four emitter array (middle). Top view of the cathode showing a two emitter array (right). 
}
\label{DiodeView}
\end{figure}

The simulation is based on the method of molecular dynamics, where every interaction between electrons in the gap is accounted for and every electron in the gap is treated as an individual particle.  The simulation algorithm is based upon three different procedures for each time-step:

\begin{subsubsection}{Emission}
For this first part, a point on the emitter is selected at random.  If the electric field at that point is favorable for emission, and electron is placed 1nm above the surface, otherwise a failure of placement is registered.  This process is repeated until no more electrons can be placed on the emitter surface.
\end{subsubsection}

\begin{subsubsection}{Advancement}
Once the new electrons have been introduced into the system at the emitter, the force on every electron in the gap, due to the external field and to Coulomb interactions, is calculated.  These force calculations are used to calculate what the electrons' positions will be in the subsequent time-step.
\end{subsubsection}

\begin{subsubsection}{Absorption and advancement of time}
Electrons that have passed the anode are eliminated from the system and the time is advanced by one-time-step.
A more detailed explanation of the simulation method can be found in \cite{pedersen2011} and \cite{jonsson2013}.
\end{subsubsection}

\begin{section}{Relative phase and the coupling parameter}
\label{Relative phase}
The relative phase is a good indicator of how well the periodic
pulses released by the two emitters are synchronized. In order to check the synchronization
in our chaotic system we use the following method \cite{kralemann2008}:
During the simulation we monitor the electrons released by each
emitter. The total signal is just the sum of the two series of pulses
produced by each emitter separately.  We will denote the signal from
the first emitter as $y_1(t)$ and the signal of the second emitter as
$y_2(t)$. We will interpret the time variable $t$ as an 
angular coordinate. By taking the Hilbert transform of the first signal we get
$y_1(t + \tau)$. If the signal is purely harmonic, $\tau$ would just be
the signal shift of $\pi /2$. $y_1(t)$ and $y_1(t + \tau)$ play the
role of dynamic conjugated variables, and they produce the phase space limit cycle of the signal
from the first emitter. From this limit cycle, the phase is easily
extracted as $\phi_1(t) = \arctan\left[\frac{y_1(t)}{y_1(t + \tau)}\right]$. We
do the same to the second signal, and from there we extract $\phi_2(t) =
\arctan\left[\frac{y_2(t)}{y_2(t + \tau)}\right]$ \cite{pikovsky2001}.

$\phi_1$ and $\phi_2$ are the phases of the two chaotic oscillators due to the two emitters.  If they are identical throughout time (or if the difference between them is a multiple of 2$\pi$), it would mean that the signals evolve in perfect synchronization.  This is generally not the case however, but if the difference between them is close to being a multiple of 2$\pi$ they are in near synchronization and add constructively.  If the difference between the phases is an odd multiple of $\pi$, they are anti-synchronized and add destructively.  In other cases they are merely unsynchronized.

It is useful to have some sort of measure of the influence of the space-charge coming from one emitter on the surface field at the center of another emitter.  We propose a {\it coupling parameter} to serve this purpose.  Consider two circular emitters located on the cathode of the same planar diode, as shown in Fig.~\ref{DiodeView}. 
The spacing between the centers of these emitters is $L$, the gap spacing is $D$, and the gap voltage is $\Phi$.  Let $E_{nm}$ be the contribution to the electric field {\it normal to the cathode} at the center of emitter $m$, due to the space-charge that originates from emitter $n$.  For example $E_{21}$ is the contribution to the electric field normal to the cathode at the center of emitter $1$, due to the space-charge that originates from emitter $2$. The coupling parameter can then be defined as $C_{21} = E_{21} / E_{11}$ or $C_{12} = E_{12} / E_{22}$.  In a symmetric system $C_{12} = C_{21}$.  Although an exact measure of this parameter is problematic it can be readily estimated.
If $\rho_1(x,y,z)$ is the charge density in the gap due to emitter $1$, and we consider the situation where emitter $2$ is turned off, then a reasonable approximation is that the charge density inside the beam emanating from the emitter is solely a function of position above the cathode, $z$, and that $\rho_1(x,y,z) = \rho_1(z) = Kz^{-2/3}$, \cite{lau2001} where $K$ is a constant.  From symmetry we anticipate that the same applies to the charge density from emitter $2$, when it is the only one emitting, i.e. $\rho_1(z) = \rho_2(z)=\rho(z)$.  Assuming this form of charge density above either emitter, when they are both emitting, it is possible to estimate the position of the center of charge, $z_c$, above each emitter
\begin{equation}
z_c = \frac{\int\limits_0^D z \rho (y) dz}{\int\limits_0^D \rho (z) dz} = \frac{D}{4}.
\end{equation}
If $Q$ is the total amount of charge present in the gap due to one emitter, then one can estimate the coupling parameter, $C$, as
\begin{equation}
C \approx \frac{Qz_c/(z_c^2+L^2)^{3/2}}{Q/z_c^2} = \frac{1}{(1+16\xi^2)^{3/2}},
\end{equation}
where $\xi=L/D$.  Although this parameter is not exact, it serves a valuable purpose in giving a quantitative measure of the effect of the space-charge from one emitter compared to that from an adjacent emitter.  Interestingly, it is solely a function of the ratio $L/D$.

\end{section}

\begin{section}{Results and analysis}
\label{Results}
We begin by examining a simple system of two emitters designed so as to keep the number of bunches in the gap larger than one, but as close to one as possible.  
To do this the radius of each emitter is set at 1 nm, the gap spacing is set at 18 nm, while the gap voltage, $\Phi$, ranges from 0-300 mV, and the spacing between emitter centers, $L$, ranges from 0-90 $\rm \mu m$ ($\xi$ ranges from 0-5).  The next step is to systematically vary $\Phi$ and $L$, and record the phase difference between absorption, at the anode, of the k-th electron from the first emitter and absorption of the k-th electron from the second emitter . 

\begin{figure} 
\begin{center}
\includegraphics [width=8.8 cm] {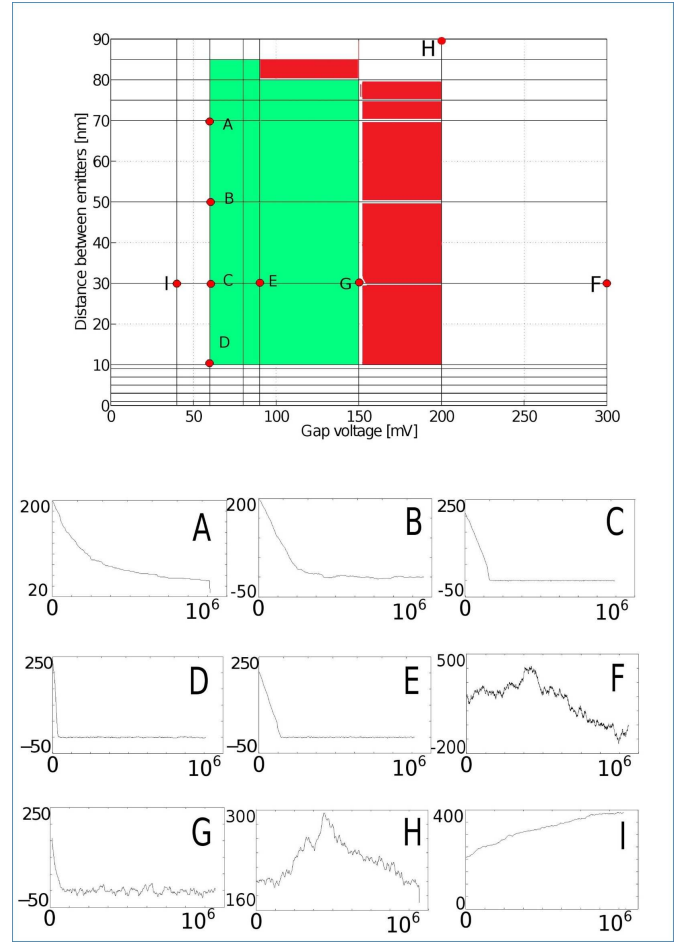}
\end{center}
\caption{
(Top) Synchronization region shaded green, transition region in red and the white region is the one where no synchronization is evident. (Bottom) The time delay diagram for each labeled point in the synchronization region and outside of it. The axes on each plot are: x-axis is the time given in time steps, y-axis is the delay given in number of emissions. In the beginning all diodes start with 200 time steps difference between each emitter.
}
\label{Syncdiagram}
\end{figure}

\begin{figure}
\begin{center}
\includegraphics [width=8.8 cm] {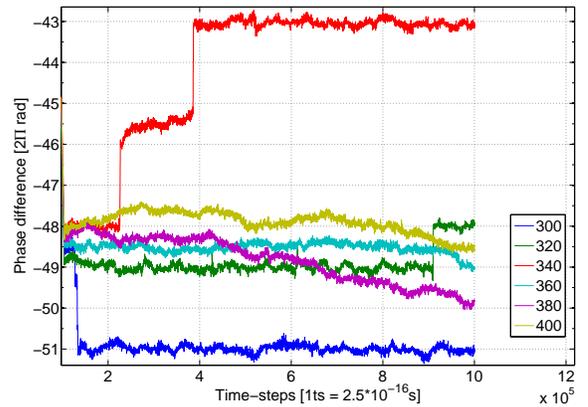}
\end{center}
\caption{The phase difference between the signals from the two
emitters.}
\label{PhaseDifference}
\end{figure}

Figure~\ref{Syncdiagram} shows results of this investigation.  
Referring to the top part of figure~\ref{Syncdiagram}, the white area shows combinations of $L$ and $D$ where no synchronization takes place or the electrons are in anti-phase, the green color indicates a region where synchronization is persistent, and the red area a transition region where synchronization drops in and out.  The bottom part of figure~\ref{Syncdiagram} shows examples of the development of the phase as a function of time for selected combinations of $L$ and $\Phi$.

The vertical boundary between the green and white region to the left of the green region is easily understood.  This is simply an area where the applied voltage is too low to support more than one electron at a time in the gap, hence one cannot speak of synchronization.  The horizontal boundary at $L$ = 10 nm comes about because the close spacing of the emitters leads to electrons released from them alternately and thus arriving at the anode in anti-phase.  The boundary at the top of the stable/transition regions is simply explained by the fact that the emitters are placed so far apart that their coupling is too weak to lead to synchronization.  Loss of synchronization because of increasing gap voltage is not as clear cut, as evidenced by the transition region to the right of the green colored area in figure~\ref{Syncdiagram}.  The reason for this loss of synchronization is that at higher gap voltage, the number of electrons present in the gap increase, and due to the small emitter size this corresponds to a rather high charge density above the emitter.  Thus mutual repulsion of the electrons disrupts the structure of the „beamlets“ and leads to degradation of synchronization.

Next we examine systems where the emitter area, gap voltage and gap spacing are all considerably larger, whereby emission occurs in electron bunches rather than as individual electrons.  This corresponds to the situation described in previous work \cite{pedersen2011}, \cite{jonsson2013}.  For subsequent simulations the gap spacing is fixed at $D$ = 500 nm, the gap voltage at $\Phi$ = 2 V, and emitter radius at $R$ = 150 nm.  The time-step used in the simulations is 0.25 fs.  All emitters are circular and flat.

When two emitters are synchronized, then their phases will change approximately together \cite{kralemann2008}, thus the phase difference between them stay approximately constant, either as an even multiple of $\pi$ for constructive synchronization, or an odd multiple of $\pi$ for destructive synchronization.  Transitions in phase difference between multiples of $\pi$, known as phase slips, may occur sporadically.  In figure~\ref{PhaseDifference} the relative phase between current from a pair of emitters is shown for different values of the emitter spacing, $L$, ranging from 300nm, when the emitters are just touching, to 400 nm.  For $L<$ 340 nm the synchronization is persistent, and mostly without phase slippage.  At $L$ = 340 nm phase slippage occurs twice, and we note that during the time interval from roughly $2 - 4 \times 10^5$ time-steps (50-200 fs) the emitters are synchronized in anti-phase leading to destructive interference.   For $L>$ 340 nm the phase difference starts to fluctuate slowly, resulting in non-persistent synchronization.

Closer examination, of the case where the two emitters are separated by $L$ = 400 nm, is instructive.  For the first 20 fs of the run, only one emitter is switched on.  At that time emission from the other emitter is allowed to commence, and for the duration of the simulation both emitters are active.  The reader is now referred to figure~\ref{PowerSpectrum}.  

\begin{figure} 
\begin{center}
\includegraphics [width=8.8 cm] {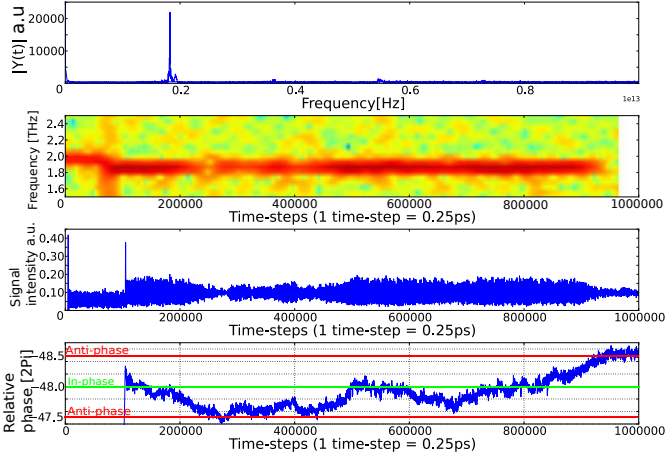}
\end{center}
\caption{
Compound picture showing (from top top bottom): a) The signal in frequency domain b) Spectrogram of the signal c) Signal in time domain d) Relative phase
(detailed explanation in text)
}
\label{PowerSpectrum}
\end{figure}

In figure~\ref{PowerSpectrum}a the frequency spectrum for the entire signal can be seen.  This includes a prominent peak at a frequency somewhat below 2 THz, and a small peak adjacent to it at a slightly higher frequency.  The smaller peak is due to the portion of the signal when only one emitter was active for the first 20 fs.  The drop in frequency is in accordance with theoretical considerations \cite{pedersen2011} and empirical evidence \cite{jonsson2013} that the bunching frequency should drop with increasing emitter area.  Further discussion of this point will be made later in this paper.

In figure~\ref{PowerSpectrum}b a spectrogram for the same signal is shown.  One may easily see the drop in frequency, and transient broadening of the spectrum, as the second emitter is turned on.  Also apparent is how the high frequency peak is eroded as the synchronization slips into anti-phase, or destructive interference, at around 80ps and towards the end of the simulation.

In figure~\ref{PowerSpectrum}c the signal is shown in time domain.  Apparent from this graph are the initial bursts of current as the emitters are turned on one after the other.  These bursts happen because, initially, the electrons are being injected into an empty (or almost empty, in the case of the second emitter) diode gap with no space-charge to inhibit them.  The strength of the synchronization shows up in the envelope of the signal oscillation and, by comparison with figure~\ref{PowerSpectrum}d, it is clear that the oscillation is strongest when the two emitters are in phase and weakest when they are in anti-phase.

We now turn our attention to the frequency of the total signal coming from multiple emitters.  As previously stated, it is known that the frequency from a single circular emitter decreases with increasing emitter area, and we wish to see if the frequency of a signal from synchronized emitters will have a frequency higher than that of a single emitter of the same area.  To test this we run a series of simulations where the frequency is measured as function of the normalized center to center spacing $\xi$.  This is done for a $1\times 2$ array and for a $2\times 2$ array.  The results are shown in figure~\ref{Frequency}.

\begin{figure} 
\begin{center}
\includegraphics [width=8.8 cm] {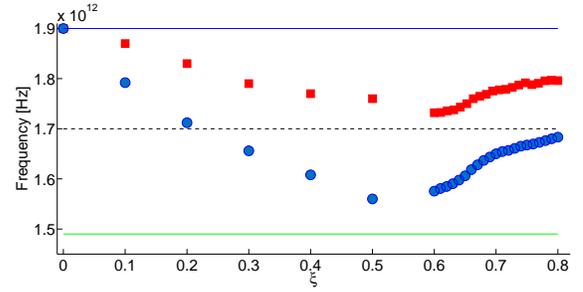}
\end{center}
\caption{Frequency dependence on the normalized center to center distance $\xi$.  The gap voltage is $\Phi$ = 2V, gap spacing is $D$ = 500 nm, and emitter radius is $R$ = 150 nm.  For $\xi<0.6$ the emitters overlap and for $\xi=0$ one emitter is completely superimposed upon the other.  Blue circles show the frequency of the signal from the $2\times 2$ array, the top solid line indicates the frequency from a single emitter of radius $R$ = 150 nm.  The middle dotted line is the frequency from a single emitter of radius $R$ = 212 nm.  The red circles show the frequency from a $1\times 2$ array and the bottom solid line shows the frequency from a single emitter of radius $R$ = 300 nm.
}
\label{Frequency}
\end{figure}

It is seen that when the emitters overlap completely, the frequency is the same as that of a single emitter of radius $R$ = 150 nm.  For the $1\times 2$ array it is clear that the frequency decreases over the interval $0<\xi<0.6$ due to the increased emission area.  When $\xi=0.6$ the emitters are barely touching, and as they are moved apart from each other they behave increasingly like independent emitters.  Thus, the frequency is at the minimum with the emitters just touching, the frequency grows as they are pulled apart.  The reader should note that the frequency of the signal from two separate emitters is always greater than the frequency from a single emitter of the same area.  A similar result can be seen for the $2\times 2$ emitter array.  The minimum frequency is obtained as the four emitters are barely touching, and the frequency from four separate emitters is always higher than that from on single emitter of the same area.  Additionally, it should be noted that the disparity in the frequency between four emitters and a single emitter of the same area, is greater than the corresponding disparity for the $1\times 2$ array.  This may indicate better coupling in the larger array.  It should also be noted that the frequency is very consistent over multiple runs using fixed parameters, hence error bars have been omitted.

A similar investigation is done for the power of the THz signal.  For each parameter combination 10 runs of the code are performed and the power is  calculated.  Figure 6 shows the normalized power of the signal from a $1\times 2$ array as a function of the normalized separation.  The power of the signal from completely overlapping emitters is one fourth of the maximum power, which occurs when the two emitters are barely touching.  It can also be seen that power increases monotonically with increasing emitter area (i.e. in the interval $0<\xi<0.6$).  As the emitters are pulled apart two items of interest can be observed. First, the total power decreases.  Second, the variance in measured power output, from the 10 different runs for each parameter set, increases.  This means that the coherence of the signal diminishes quite rapidly with separation beyond touching.

\begin{figure}
\begin{center}
\includegraphics [width=9 cm] {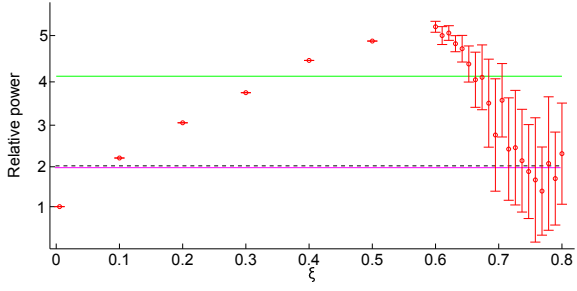}
\end{center}
\caption{(Two emitters) After performing the Fourier transform of the
total signal, we take a region centered around the main frequency peak
and take the integral under the curve. This gives us the power of the
signal in this region. The power of the signal in this region is actually
the power that we would be extracting from the device. As the emitters
are pulled further apart the power of the total signal drops. The green
full line represents the power of one emitter. The green dashed shows
twice the power of one emitter and this is coincidentally the same as
the average power of two uncorrelated signals.
} 
\label{PowerFourier}
\end{figure}

\begin{figure}
\begin{center}
\includegraphics [width=9 cm] {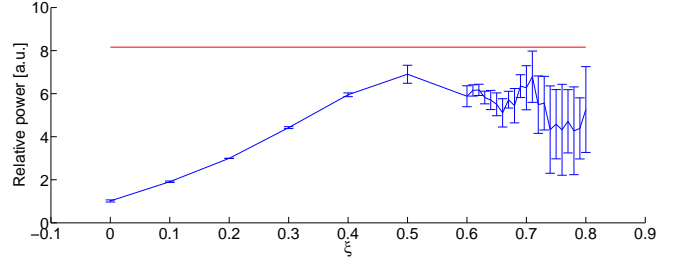}
\end{center}
\caption{(Two emitters) Power of $2\times 2$ array. The red
full line represents the power of one emitter with identical area as four single emitters of R=150 nm.  
} 
\label{PowerFourier2x2}
\end{figure}

Let us now consider $N$ sinusoidal signals of equal magnitude and phase, but varying frequency: $i_k = sin(\omega_t+f_k)$ for $k = 1$ to $N$. These signals are added together to form a compound signal $i_s = \sum\limits_{k=1}^N i_k $. If we let $P$ denote the time-averaged power of signal $i_k$ (for $k = 1, 2, ..., N$) and $P_s$ denote the time-averaged power of the compound signal, then one can readily see that 
\begin{equation}
P_s = \frac{\omega}{2 \pi} \int_0^{2 \pi/\omega} i_s^2 dt = NP +2P \sum\limits_{r \neq s} cos(\phi_r -\phi_s), 
\end{equation}
where the sum is taken over all $(N^2-N)/2$ possible combinations of $r \neq s$ with $r$ and $s$ taking integer values from 1 to $N$. The highest achievable value for $P_s$, if all signals are in phase, is $N^2P$ and the power averaged over a uniform distribution of oscillator phases, $f_k$, is $N P$. For the special case of $N = 2$, we see that $P_s$ can range from 0 to 4$P$ with a value of 2$P$ when averaged over phase difference between the signals.  From figure~\ref{PowerFourier} it can be seen that the average power from two separate emitters tends to cluster around twice the power from a single emitter, and that the variance also increases as the separation grows.  Additionally a signal, $g(t)$, from one emitter is generated  a time shifted signal, $g(t-\tau)$, produced from it.  
From this a compound signal $G(t;\tau) = g(t) + g(t-\tau)$ is constructed.  Hence, the power carried by $G(t;\tau)$ can be calculated and averaged over $\tau$.  This average power is shown open circles in figure~\ref{PowerFourier}, and matches the expected value of twice the power from one emitter quite well.

Also shown in figure~\ref{PowerFourier} is the  power from a single emitter of the same area as the two seperate emitters.  One may observe that this power is slightly less than the peak power obtained with two emitters barely touching.

Figure~\ref{PowerFourier2x2} shows how the power is affected by the emitter separation, $\xi$, in a $2\times2$ array of emitters. Each emitter has a radius of 150nm. It is apparent that the $2\times2$ array shares similar characteristics with the $1\times2$ array with the relative power increasing as the overlap decreases. However, in this case the peak power occurs at $\xi = 0.5$ rather than $\xi = 0.6$ as before, and does not match the power output of a circular emitter of 300nm radius. As in figure~\ref{PowerFourier} we can see that for $x > 0.7$ phase synchronization seems to vanish as the power distribution is more similar to what would be expected from four independent emitters.

\end{section}

\begin{section}{Discussion and summary}
\label{Discussion}
Previous studies have indicated that, under certain conditions, space-charge limited current from an emitter of limited area in a planar microdiode, will spontaneously form bunches so that the beam current is modulated with a frequency in the THz regime.  The frequency is dependent on the applied field with which it grows according to a power law, and also upon the size of the emitter, with the frequency decreasing as the emitter area increases \cite{jonsson2013}.  In this work it is shown that the current from individual emitters can synchronize via Coulomb interaction, if the emitters are not too far apart.  

For a given applied field and total emitter area, the frequency of the synchronized system depends on the spacing between the emitters, but is greater than the frequency from a single emitter in all cases.  The power coming from two circular emitters, that are barely touching, is shown to slightly exceed the power from a single emitter of the same total area.  However, the average power from the emitter pair drops off rapidly as the distance between them is increased, settling around a value that corresponds to the average power from the sum of two sinusoidal currents with the same frequency but randomized phase difference.  This indicates that frequency locking is much more persistent than phase locking.

Similar results are observed for the $2\times2$ array, although the gain in power is not the same as for the $1\times2$ array. On the other hand the $2\times2$ array shows better frequency characteristics than the $1\times2$ array in the sense that it is proportionally higher than the frequency from a circular emitter of the same area as the array.

We also show that for a simple system consisting of electrons coming from two point emitters a certain parameter range, in terms of applied field and spacing between emitters, leads to synchronization.  In other words, a „sweet-spot“ for synchronization exists.

Our simulations show that it is possible to extract more power, at a higher frequency, from an array of emitters than would be possible from a single emitter of the same total area. To examine how this effect may be optimized, and how it applies to large arrays, is beyond the scope of this paper but will be examined in future work both via simulation and experiment. Examination of stronger coupling mechanisms than the simple Coulomb interaction also merit further investigation.

\end{section}


%


\section*{Acknowledgment}

The authors would like to thank Arkady Pikovsky and Michael Rosenblum from Potsdam University for helpful discussion.  This work was supported by the Icelandic Research Fund grant number 120009021.

\ifCLASSOPTIONcaptionsoff
  \newpage
\fi


\begin{thebibliography}{1}

\bibitem{siegel2002a}
P.H. Siegel, “Terahertz technology“, IEEE Trans. Microwave Theory Techn., vol. 50, pp. 910-928, 2002.

\bibitem{siegel2002b}
P.H. Siegel, “Terahertz technology in biology and medicine“, IEEE Trans. Microwave Theory Techn., vol. 52, pp. 2438-2447, 2002.

\bibitem{booske2008}
J.H. Booske, “Plasma physics and related challenges of millimeter-Wwve-to-terahertz and high power microwave generation“, Phys. Plasmas, vol. 15, pp. 1-16, 2008.

\bibitem{booske2011}
J.H. Booske, R.J. Dobbs, C.D. Joye, C.L. Kory, G.R. Neil, G.S. Park, J. Park, and R.J. Temkin, “Vacuum electronic high power terahertz sources“, IEEE Trans. Terahertz Sci. Techn., vol. 1, pp. 54-75, 2011.

\bibitem{scalari2009}
G. Scalari, C. Walther, M.  Fischer, R. Terazzi, H. Beere, D. Ritchie, and J. Faist, “THz and sub-THz quantum cascade lasers“, Laser \& Photon Rev., vol. 3, pp. 45-66, 2009.

\bibitem{tonouchi2007}
M. Tonouchi, “Cutting-edge terahertz technology“, Nature Photonics, vol. 1, pp. 97-105, 2007.

\bibitem{williams2007}
B.S. Williams, “Terahertz quantum-cascade lasers“, Nature Photonics, vol. 1, pp. 517-525, 2007.

\bibitem{jepsen2011}
P.U. Jepsen, D.G. Cooke, and M. Koch, “Terahertz spectroscopy and imaging –Modern techniques and applications“, Laser \& Photon Rev., vol. 5, pp. 124-166, 2011.

\bibitem{federici2010}
J. Federici and L. Moeller, “Review of terahertz and subterahertz wireless communications“, J. Appl. Phys., vol. 107, 11101, 2010.

\bibitem{booske2005}
R.J Booske, J.H. Booske, N.C. Luhmann, and G.S. Nusinovich, Modern Microwave and Millimeter-Wave Power Electronics, Piscataway, NJ: IEEE, 2005.

\bibitem{freund1999}
H.P. Freund and G.R. Neil, “Free-electron lasers: Vacuum electronic generators of coherent radiation“, Proc. IEEE, vol. 87, pp. 782-803, 1999.

\bibitem{nusinovich2004}
G.S. Nusinovich, Introduction to the Physics of Gyrotrons, Baltimore, MD.: John Hopkins Univ. Press, 2004.

\bibitem{symons1998}
R.S. Symons, “Tubes:  Still vital after all these years“, IEEE Spectrum, vol.35, pp. 52-63, 1998.

\bibitem{shin2009}
Y.M. Shin, L.R. Barnett, D. Gamzina, N.C. Luhmann, M. Field, and R. Borwick, “Terahertz vacuum electronic circuits fabricated by UV lithographic molding and deep reactive ion etching“, Appl. Phys. Lett., vol.95, 181505, 2009.

\bibitem{zhu2001}
W. Zhu, Vacuum Microelectronics, New York NY: Wiley, 2001.

\bibitem{bhattarjee2004}
S. Bhattarjee, J.H. Booske, C.L. Kory, et al., “Folded waveguide traveling-wave tube sources for terahertz radiation“, IEEE Trans. Plasma Sci., vol. 32, pp. 1002-1014, 2004.

\bibitem{ives2004}
R. Ives, “Microfabrication of high-frequency vacuum electron devices“, IEEE Trans. Plasma Sci., vol. 32, pp. 1277-1291, 2004.

\bibitem{pedersen2011}
A. Pedersen, A. Manolescu, and A. Valfells, “Space-charge modulation in vacuum microdiodes at THz frequencis“, Phys. Rev. Lett., vol. 104, 175002, 2011.

\bibitem{birdsall1966}
C. Birdsall and W. Bridges, Electron Dynamics of Diode Regions, New York NY: Academic Press, 1966.

\bibitem{lau2001}
Y.Y. Lau, “Simple theory for the two-dimensional Child-Langmuir law“, Phys. Rev. Lett., vol. 87, 278301, 2001.

\bibitem{zhu2011}
Y.B. Zhu, “Child-Langmuir law in the Coulomb blockade regime“, Appl. Phys. Lett., vol. 98, 051502, 2011.

\bibitem{valfells2002}
A. Valfells, D.W. Feldman, M. Virgo, P.G. O‘Shea, and Y.Y. Lau, “Effects of pulse-length and emitter area on virtual cathode formation in electron guns“, Phys. Plasmas, vol. 9, pp. 2377-2382, 2002.

\bibitem{luginsland2002}
J.W. Luginsland, Y.Y. Lau, R.J. Umstattd, and J.J. Watrous, “Beyond the Child-Langmuir law: A review of recent results on multidimensional space-charge-limited flow“, Phys. Plasmas, vol. 9, pp. 2371-2376, 2002.

\bibitem{caflisch2012}
R.E. Caflisch and M.S. Rosin, “Beyond the Child-Langmuir limit“, Phys. Rev. E, vol.85, 056408, 2012.

\bibitem{jonsson2013}
P. Jonsson, M. Ilkov, A. Manolescu, A. Pedersen, and A. Valfells, “Tunability of the terahertz space-charge modulation in a vacuum microdiode“, Phys. Plasmas, vol.21, 023107, 2013.

\bibitem{zhu2013}
Y.B. Zhu, P. Zhang, A. Valfells, L.K. Ang, and Y.Y. Lau, “Novel scaling laws for the Langmuir-Blodgett solutions in cylindrical and spherical diodes“, Phys. Rev. Lett., vol. 110, 265007, 2013. 

\bibitem{pikovsky2001}
A. Pikovsky, M. Rosenblum, and J. Kurths, Synchronization: A Universal Concept in Nonlinear Sciences, Cambridge UK: Cambridge Univ. Press, 2001.

\bibitem{kralemann2008} 
B. Kralemann, L. Cimpoieriu, M. Rosenblum, A. Pikovsky, “Phase dynamics of coupled oscillators reconstructed from data“, Phys. Rev. E, vol. 77, 066205, 2008. 


\end{thebibliography}
\end{document}